# Efficient, High-purity, Robust Sound Frequency Conversion with a Linear Metasurface

Chengbo Hu[1*], Wei Wang[1*], Jincheng Ni[2*], Yujiang Ding[1], Jingkai Weng[1], Bin Liang[1†], Cheng-Wei Qiu[2†] and Jianchun Cheng[1†]

[1]*Key Laboratory of Modern Acoustics, MOE, Institute of Acoustics, Department of Physics, Collaborative Innovation Center of Advanced Microstructures, Nanjing University, Nanjing* 210093*, P. R. China*

[2]*Department of Electrical and Computer Engineering, National University of Singapore, 117583 Singapore, Singapore.*

[*]These three authors contributed to this work.

[†]To whom correspondence may be addressed. Email: liangbin@nju.edu.cn (B. L.), chengwei.qiu@nus.edu.sg (C. Q.), or jccheng@nju.edu.cn (J. C.).

**Abstract**
The intrinsic limitation of the material nonlinearity inevitably results in the poor mode purity, conversion efficiency and real-time reconfigurability of the generated harmonic waves, both in optics and acoustics. Rotational Doppler effect provides us an intuitive paradigm to shifting the frequency in a linear system, which needs to be facilitated by a spiraling phase change upon the wave propagation. Here we numerically and experimentally present a rotating linear vortex metasurface and achieve close-to-unity mode purity (above 95%) and conversion efficiency (above 65%) in audible sound frequency as low as 3000 Hz. The topological charge of the transmitted sound is almost immune from the rotational speed and transmissivity, demonstrating the mechanical robustness and stability in adjusting the high-performance frequency conversion in situ. These features enable us to cascade multiple vortex metasurfaces to further enlarge and diversify the extent of sound frequency conversion, which are experimentally verified. Our strategy takes a step further towards the freewheeling sound manipulation at acoustic frequency domain, and may have far-researching impacts in various acoustic communications, signal processing, and contactless detection.

**Significance**
Frequency conversion is of fundamental interest in wave physics and practical significance for diverse applications. However, high-efficiency conversion of low-frequency waves cannot be achieved with the current mechanisms based on nonlinear medium or up/down-conversion materials. Here, we theoretically propose and experimentally demonstrate a mechanism that bypasses these limitations with a rotating metasurface. In particular, the resulting frequency-conversion device high-efficiently and precisely converts the incident wave of arbitrary wavefront into the desired acoustic vortex mode and enables device cascading to further amplification of frequency shift. Our strategy with unprecedented advantages of close-to-unity mode purity, high conversion efficiency, wavefront robustness and cascadability opens a route to freewheeling sound manipulation at frequency domain, with important applications in acoustics and many related fields.

**Introduction**
Frequency conversion is of fundamental interest in wave physics and conventionally realized by utilizing nonlinear, up/down-conversion materials in optics. The most intuitive way of frequency conversion for light is to use the nonlinear process in a strong laser electric field, in which the energy of fundamental wave will be partially converted to high-order harmonic waves (1). By Stokes/anti-Stokes process, photon down/up-conversion materials can emit a lower/higher energy photon than the absorbed photons (2, 3). Despite a rapid progress in the research field, wide applications of frequency conversion in optics are strongly limited because of their overall low efficiencies. In acoustics, inhomogeneity of medium such as pulsating bubbles can significantly enhance the equivalent nonlinearity parameter (4, 5). However, the intrinsic dependence of frequency and driving amplitude limits the efficiency of nonlinear frequency conversion and makes it impossible to produce only one harmonic wave for low-frequency and weak-amplitude sound. In linear regime, frequency conversion can also be observed on the basis of Doppler effect, which stems from the relative motion of wave source and receiver. This can be classified into two categories: 1) the translational Doppler effect which is caused by the translational motion of receiver or wave source and is extensively used for detecting an object's moving velocity with light or acoustic wave (such as in speed detection), and 2) the rotational Doppler effect which happens when there is relative rotational motion of spiral wave front and receiver and leads to frequency shift that can be used to detect the rotation speed of objects. For electromagnetic waves, it was reported that the rotational Doppler effect arises from the photonic orbital angular momentum (OAM) by vortex beams (6). The vortex beams have a helical phase structure of $\exp(il\phi)$ proportional to the azimuthal angle $\phi$ with an unbounded integer $l$ referring to the topological charge (7-13). The OAM of light has grown into a significant research field, giving rise to many developments in optical sensing, micromanipulation, micro/nanofabrication, imaging and microscopy, and optical communications (14-18). In particular, vortex generator with tunable topological charges is a critical step in the realization of OAM modulation and multiplexing both in optical and acoustic communications (19-21). Generally, the vortex generators are designed for specific frequencies, implying that phase distortion or mode crosstalk would be inevitable after the operating frequency is finely altered. On the other hand, the OAM-frequency division multiplexing for multidimensional high-capacity information processing requires stable OAM modes at different frequencies, which is not yet accessible by static devices. The rotational Doppler frequency shift for photonic OAM can be given as $\Delta\omega = l\Omega$ in the scheme with the vortex beam detected by a transducer rotating around the beam axis (22-31). Also, the backscattering of light carrying OAM on a rough rotating surface has been shown to undergo a rotational Doppler frequency shift (31).

Due to the longitudinal nature of acoustic waves lacking spin angular momentum, airborne sound carrying OAM can be generated by using active and passive devices (7, 32, 33). In particular, the rotational Doppler frequency shift of sound has been observed acoustically, but needs an incident acoustic vortex beam and a spinning detector which would be difficult to implement in practice (34). To date, a high-efficiency and high-purity frequency conversion of low-frequency sound in static medium still remains challenging.

In the present work, we theoretically propose and experimentally demonstrate a mechanism for overcoming the above limitations by a rotation system and a high-transmittance acoustic metasurface that can load transmitted acoustic waves with an additional OAM. The phase-adjustable and high-transmittance units are periodically arranged along the azimuthal direction, forming a spiral phase distribution on the transmission surface with $l$-order OAM with the plane wave incident at working frequency $f_0$. A rotation system drives the metasurface to rotate steadily around the axis of symmetry at an angular velocity of $\Omega$. In this case, the phase distribution on the transmission surface leads to the rotational Doppler effect and the transmitted wave can be converted to a new frequency expressed as $f_0 + l\Omega/2\pi$. Meanwhile, the high transmittance of the metasurface maintains in the rotation and keeps the conversion effective. We also demonstrate the robustness of the incident wavefront experimentally, meaning that high efficiency and frequency shift can be maintained regardless of the incident wavefront. This enables cascading of the frequency convertor for further enhancing the frequency shifts, which is also verified experimentally on frequency conversion and efficiency.

**Results**

Figure 1A and B illustrates a schematic of the frequency convertor based on the rotational Doppler effect. By a plane wave incidence at the working frequency, a transmitted acoustic vortex beam with an azimuthal phase distribution expressed as $l\phi$ can be produced by the metasurface with a specific topological charge $l$. The metasurface can twist the sound plane wave into a helical wavefront by adding phase factor of $e^{il\phi}$, only changing the orientation of sound wavevector (Fig. 1A). Note that the frequency of the generated acoustic vortex is equal to the fundamental frequency. When we rotate the metasurface at a frequency $\Omega$, the helical wavefront is further twisted or released with a phase factor of $e^{il\Omega t}$ in the original propagating distance (Fig. 1B). In this scenario, both the orientation and magnitude of sound wavevectors are changed, implying an acoustic frequency shift of $\Delta f = l\Omega/2\pi$. For a clear description, a rest frame is defined as $(r, \phi, z)$ in cylindrical coordinate, where $r, \phi$ are the radial and azimuthal direction of the metasurface, and $z$ is identical to the incident direction of the acoustic wave, as shown in Fig. 1C. When relative motion between the acoustic wave and the metasurface appears in the azimuthal direction, the additional phase shift ($l\phi$) carried by the metasurface can generate the frequency shift ($l\Omega/2\pi$) of the transmitted sound wave through the rotation of the metasurface. Assuming that a plane wave is incident vertically on a rotating metasurface, the azimuthal phase distribution of the transmitted wave can be described as $l\phi'$ in the rotational frame. Here the rotational frame referenced to the rest frame is described as $(r', \phi', z')$ in cylindrical coordinates, where $r' = r$, $\phi' = \phi - \Omega t$ and $z' = z$. The plane wave can be expressed as $p_i = e^{i(\omega t - k_z z)}$ and $p_i' = Ae^{i(\omega t - k_{z'} z')}$ in the rest and the rotational frame, respectively, where $A$ is the amplitude of the plane wave. Considering the high transmittance of the metasurface, we adopt an ideal situation without scattering and loss that the transfer function of the metasurface can be expressed as $e^{-ik_z d}e^{il\phi}$, where $d$ is the thickness of the metasurface. The expression of the transmitted wave in the rotational frame can be derived as:

$$p_t'|_{z'=d} = p_i'|_{z'=0}e^{-ik_{z'}d}e^{il\phi'} = Ae^{i(\omega t - k_z d)}e^{il\phi'} \qquad (1)$$

where $z' = 0$ and $z' = d$ are the incident and transmission planes of the metasurface respectively. The transmitted wave in the rest frame can be expressed as:

$$p_t|_{z=d} = p_t'|_{z'=d} = Ae^{i(\omega - l\Omega)t}e^{i(l\phi - k_z d)} \qquad (2)$$

Hence, one readily derives the frequency of transmitted wave probed by a static microphone behind the rotating metasurface can be derived as $f' = \frac{\omega - l\Omega}{2\pi}$, and a high efficiency frequency convertor can be obtained by a metasurface with a high transmittance. From Eqs. (1) and (2) one observes that the amplitude $A$ is independent of the frequency conversion, which implies the robustness of our mechanism to the incident wavefront as will be verified later.

Series numerical simulations have been carried out to demonstrate the performance of the designed Doppler frequency convertor. Figure 2 demonstrates the simulated Fourier spectra of the transmitted wave behind 2 and 3-order metasurface with different rotating rates, respectively. The working frequency of the 2$^{nd}$ and 3$^{rd}$ -order metasurface are 3000 and 3100Hz, respectively. The Fourier spectra of the transmitted acoustic waves show that the numerical results are in good agreement with the theoretical prediction. With the increase and decrease of rotation rates, the frequency of the transmitted waves will shift according to $f_0 + l\Omega/2\pi$, which also agrees well with the experimental results in Fig. 1D. The amplitude of the main peaks in all spectra is similar to that of static case ($\Omega = 0$), which shows that a high-efficiency conversion of the frequency can be easily achieved by a rotating metasurface that can produce a specific order acoustic vortex. Insets in Figures 2A and B demonstrate the amplitude and phase distribution of the transmitted acoustic waves at the cross-section 5.7 cm behind the metasurface with the rotational frequency equal to 0 Hz and 50 Hz, respectively. In the transmitted waves, only a small part of the acoustic energy remains in the fundamental component, and most of energy is converted into the harmonic component. The ratios of harmonic-component intensity in transmitted wave to the total intensity of incident wave and transmitted wave are defined as the efficiency and purity of frequency conversion, respectively, which is above 65% and 95% in numerical simulation. Noting that the topological charge of the harmonic component in the transmitted wave is the same as that of the metasurface and will not be affected by the rotating speed, but the fundamental component still maintains the original topological charge of the incident wave. This means that only acoustic waves with additional phase modulation from metasurface could result in frequency conversion.

Next, we performed experiments to verify the results of the theoretical prediction and numerical simulations. Figure 3A demonstrates the experimental setup for frequency conversion produced by an acoustic metasurface with different topological charge $l$ and rotational frequency $\Omega/2\pi$. Also, a measured phase distribution (Fig. 1D) at the cut-plane perpendicular to the axial direction 20 cm behind the metasurface illustrates an acoustic vortex ($l = 2$) with 60 Hz frequency shift. These all agree well with the results from the numerical simulation except for slight reduction in the efficiency of the frequency conversion which results from the influence of the scattered wave produced by the bearing bracket and other essential parts in the rotating transmission system, and indicate that the rotation of the metasurface does not destroy the vortex wavefront that the metasurface should have produced in static media. Also, the fundamental frequency component in the transmitted waves mainly results from the leakage of the fundamental waves caused by the slit between the waveguide and the rotating metasurface. Nevertheless, according to the numerical simulations, above 95% purity can be achieved in the frequency convertor, which indicates that the conversion efficiency can be nearly perfect in our ideal model given the metasurface only changes the propagation phase while keeping a near-unity transmission efficiency. The measured efficiency is slightly lower than simulation due to the imperfect impedance match between air and the experimental samples and the scattering from the necessary fixing and transmission accessories.

Figures 3B and C demonstrate the Fourier spectra of the transmitted wave behind $2^{nd}$ and $3^{rd}$ order metasurfaces with different rotating rates, respectively. The working frequencies of the $2^{nd}$ and $3^{rd}$ order metasurfaces are 3000 Hz and 2900 Hz, respectively. The Fourier spectra of the transmitted acoustic waves show that the theoretical prediction is in good agreement with the experimental results. In complete agreement with the results of the numerical simulations, the amplitude of the main peaks in all spectra is similar to that of static case ($\Omega = 0$), which indicates that the experimental samples made from photopolymer can easily maintain high-purity and high-efficiency frequency conversion through rotating mechanisms. The peaks referring to fundamental frequency with small amplitude remain on the spectrum, which indicates that the energy of the fundamental waves almost completely converts to a new frequency through a simple rotating system. As the theoretical prediction mentioned, the frequency conversion is based on perfect spiral phase distribution or OAM mode, so that the harmonic component in the transmitted waves has perfect spiral phase distribution of OAM, which is verified numerically and experimentally in Figs. 2 and 1D. However, the fundamental component cannot obtain OAM from the metasurface and keeps the original wave front without spiral phase distribution, as shown in the insets of Fig. 2. Furthermore, owing to the additional spiral phase distribution of $e^{il\phi}$ and efficiency of the frequency conversion, the convertor is potential to serve as a frequency-tunable OAM generator and more versatile wave manipulation, such as frequency modulator.

This efficient frequency convertor also has high robustness to incident wavefront. As a representative example, we use obstacles, such as plank, to block half of the frequency convertor, in which case the incident wavefront is no longer a plane wave and becomes a diffracted wave on the rim of acoustically-rigid plate, as shown in Fig. 4A. Behind the metasurface, a perfect acoustic vortex cannot be formed because of the absence of the acoustic waves transmitted from half of the blocked area, but the efficient frequency shift is still verified experimentally, as shown in Fig. 4B. Also, the frequency convertor still works under the incidence of the acoustic vortex waves. A static metasurface is utilized to produce specific order acoustic vortex wave, which further radiates on a rotating metasurface, as shown in Fig. 4C. The Fourier spectra of the transmitted wave under different rotating rates shown in Fig. 4D prove the efficient conversion persists as we modify the rotating rates which only changes the shift of frequency, indicating that an efficient cascade scheme of the frequency convertor can be realized to amplify the frequency shift.

Figure 5A explains the mechanism of frequency conversion in such a system consisting of two cascading metasurfaces with topological charges chosen as $l_1$ and $l_2$ respectively, and what happens when an incident wave passes through these two layers. Each metasurface with its respective topological charge and rotating speed will give rise to an extra frequency shift $\Delta f_i$ to the transmitted wave passing through it and the frequency of the output beam will be converted into $f_0 + \sum_i^N \Delta f_i$, where $N = 2$ is the number of the metasurfaces and $\Delta f_i = \Omega_i l_i / 2\pi$ with the subscript $i$ correspond to the sequence the metasurfaces in the cascade scheme. Illuminated by the incident acoustic wave with a frequency of $f_0$ equal to the working frequency of the first metasurface $l_1$, the first frequency convertor twists the wave front of the incident plane wave into a vortex with topological charge equal to $l_1$ and converts the frequency of the wave to $f_0 + \Delta f_1$. When the new frequency meets with the working frequency of the next metasurface $l_2$, a further frequency conversion happens with the further frequency shift and additional topological charge equal to $\Delta f_2$ and $l_2$. As shown in Fig. 5E, theoretical predictions and experimental measurements are in good agreement for different frequency manipulations by cascade scheme of the frequency convertor. The frequency of the acoustic wave passing through the first frequency convertor can be further manipulated in frequency domain. The topological charge and rotational angular velocity are tunable, which endows a lot of freedom for frequency manipulation. Figure 5C verifies the efficiency of the cascade scheme with different angular velocities of the second metasurface $l_2$. There is almost no component for fundamental frequency in the spectrum of the transmitted waves in Fig. 5C, indicating a high purity conversion still maintained in the cascade scheme. Also, all the amplitude of the shifted frequencies at different angular velocities are similar that of the static one ($\Omega_2 = 0$ rad/$s$), verifying the high efficiency of the cascading scheme. The high efficiency of the cascade scheme is also illustrated in Fig. 5D despite of the incomplete matching between frequency of the incident wave and the working frequency of the next cascade frequency convertor resulting from the errors and limitations brought by the transmission system.

**Discussion and Conclusions**
In summary, we present here the theoretical, numerical, and experimental work of the design of a frequency convertor and its cascade scheme for efficiently converting the frequency of the acoustic wave to a new frequency in rest frame. A rotating metasurface implementation is presented by a metasurface based on hybrid resonant unit cells in a waveguide driven by a servo motor. The designed frequency convertor and its cascade scheme are new approaches of manipulation of acoustic waves in frequency domain. After passing through the frequency convertor, the frequency of the incident acoustic waves can be efficiently converted to a new frequency with a specific additional OAM by a frequency shift which depends on the topological charge and the rotational angular velocity of the metasurface in the frequency convertor. Owing to the high efficiency, this frequency conversion mechanism can extend to a cascade scheme by cascading several frequency convertors to achieve a higher or lower frequency than the fundamental frequency. The cascade scheme has been verified efficiently by cascading two frequency convertors in experiments. In addition to the simple design, high efficiency and frequency shift capability, we still need to stress several important advantages of our mechanism: (1) It offers the flexibility of freely designing the unit cells in the rotating metasurface, given that the desired phase distribution as a *l*-order acoustic vortex at the transmitted plane can be achieved for the incident plane wave at working frequency. This significantly enhances the versatility and application potential of the frequency convertor (2) Robustness of the frequency convertor is verified experimentally which is important for its application in diverse scenarios with unpredictable incident wavefronts. (3) The capability of efficient amplifying the frequency shift via a cascade scheme provide a platform for investigating many phenomena resulting from rotational Doppler effect, such as reversal of orbital angular momentum arising from an extreme Doppler shift. (4) Such a rotating metasurface can efficiently generate acoustic vortex beams with stable topological charges at varying frequency. The OAM generators with tunable frequency will significantly impact the future acoustic communications. (5) The precise correspondence between frequency shift and rotating rates enables our design to measure the actual rotating rates of a rotating object. (6) Our design can be equivalent to a time-varying metamaterial, offering the possibility of finding more extensive applications in noise control, biomedical applications and ultrasound control, etc.

**Methods**
Figures 3A and 5B demonstrate the experimental systems for frequency convertor and the cascade scheme. A metasurface made of photopolymer is driven by a servo motor via a transmission shaft and tooth synchronous belt. The bearing system fixed to the vibration isolation platform is used to ensure smooth rotation of the shaft without axial slip. Acoustic waves generated from a compression driver with a horn are incident on the rotating metasurface, which is covered by a waveguide to better mimic a plane wave and avoid the influence of the diffraction of the incident waves. The wedge made of sound-absorbing foam is placed at the end of waveguide to eliminate the undesired reflection. The radius of the metasurface used in the frequency convertor is 9.15 cm which is about 0.8 wavelength of the working frequency. All the metasurfaces are fabricated by 3D printing technology and fixed with the shaft through flanges. The design of the metasurfaces is demonstrated in Supplementary Materials. There are some unavoidable vibrations of the rotating metasurfaces caused by centrifugal action in the experiments so that the radius of the waveguide, set as 9.5 cm, needs to be slightly larger than that of the metasurfaces. The 0.25-in. Brüel and Kjær type-4961 microphones were used to measure the sound pressure distribution behind the rotating metasurface in the waveguide. An additional microphone of the same type is used as the reference microphone in the measurement of the phase distribution of transmitted waves. Servo motors provide the rotating power for the metasurfaces through the gears and belts with the speed ratio of 1:1. In order to eliminate the error between the actual rotating speed and the number displayed on the servo motor controller, the speed in all experiments was calibrated by a tachometer whose error is less than 1/60 Hz.

**Acknowledgments**
This work was supported by the National Key R&D Program of China (Grant No.

2017YFA0303700), the National Natural Science Foundation of China (Grant Nos. 11634006, 11704284, 11374157, and 81127901), the National Research Foundation, Prime Minister's Office, Singapore under Competitive Research Program Award NRF-CRP22-2019-0006, the Innovation Special Zone of National Defense Science and Technology, High-Performance Computing Center of Collaborative Innovation Center of Advanced Microstructures, and a project funded by the Priority Academic Program Development of Jiangsu Higher Education Institutions.

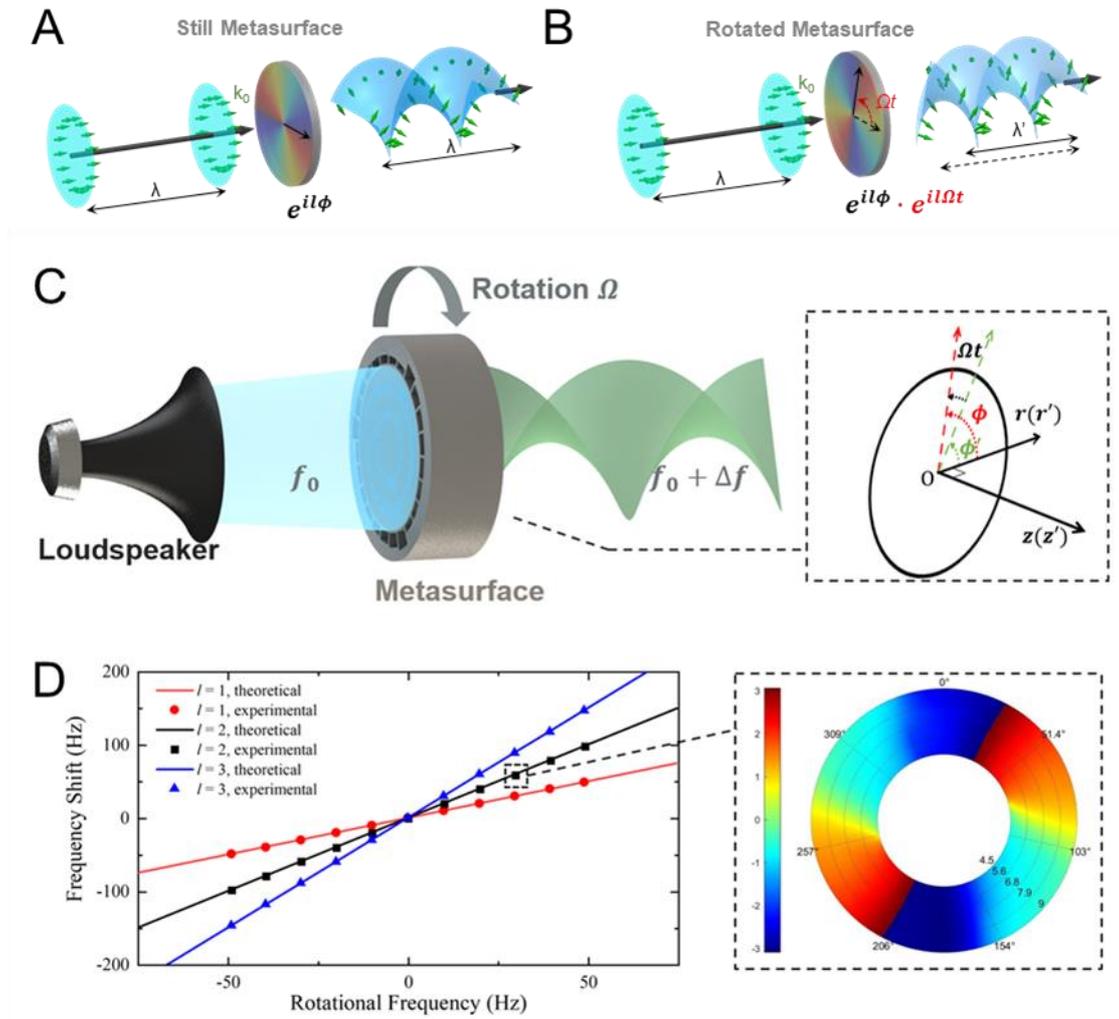

**Figure 1** | (A) Schematic of generation of acoustic vortex with topological charge $l = 2$ by a still metasurface, where the sound wave is tuned to helical wavefront. (B) The rotation of metasurface can twist the wavefront, resulting in an acoustic frequency shift. (C) Schematic of the frequency convertor: frequency of the acoustic wave will be converted to a new frequency after passing through a specifically designed rotating metasurface. Inset: schematic of coordinate transformation. (D) Theoretical and perimental results of the frequency shift with different topological charge *l* and rotational frequency $\Omega/2\pi$. Inset: phase distribution of the harmonic frequency at the cross section 5.7 cm behind the metasurface with $l = 2$ and $\Omega/2\pi = 30$ Hz.

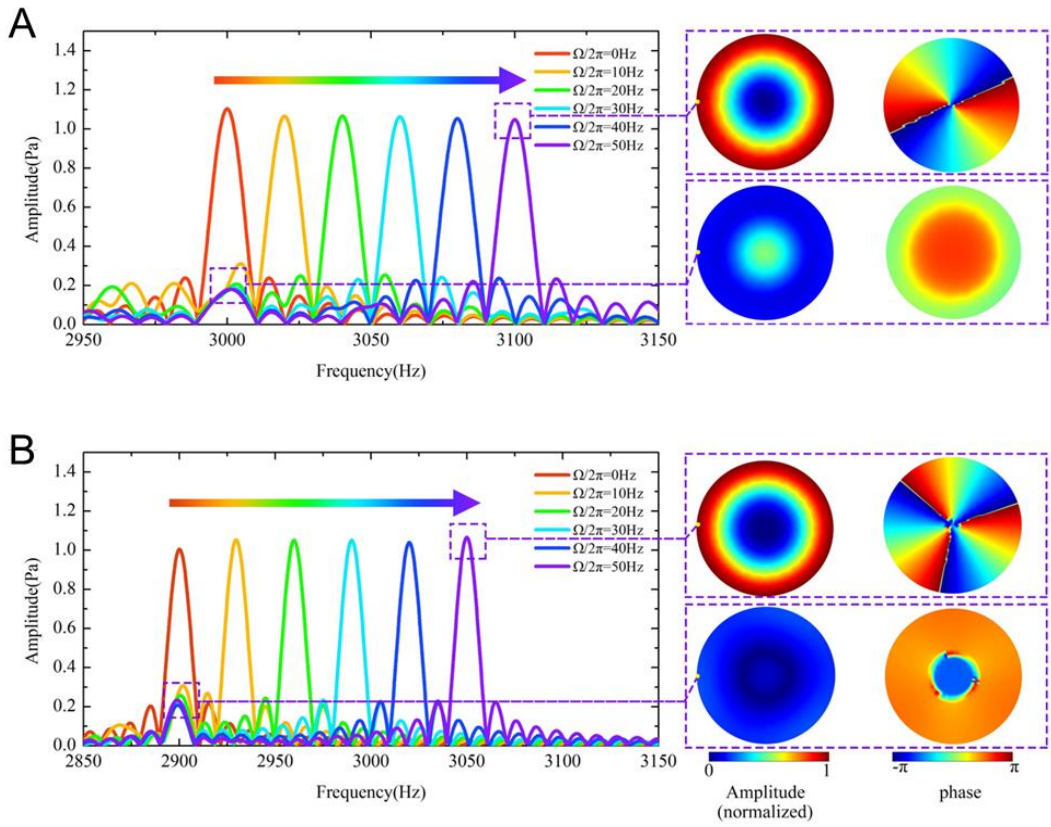

**Figure 2** | (A) Numerical results of the frequency conversion under different rotating frequency with topological charge $l$ equal to (A): 2 and (B): 3. Inset: amplitude and phase distribution of the harmonic and fundamental frequency at the cross section 5.7 cm behind the metasurface with $l = 2$ and 3, $\Omega/2\pi = 50$ Hz.

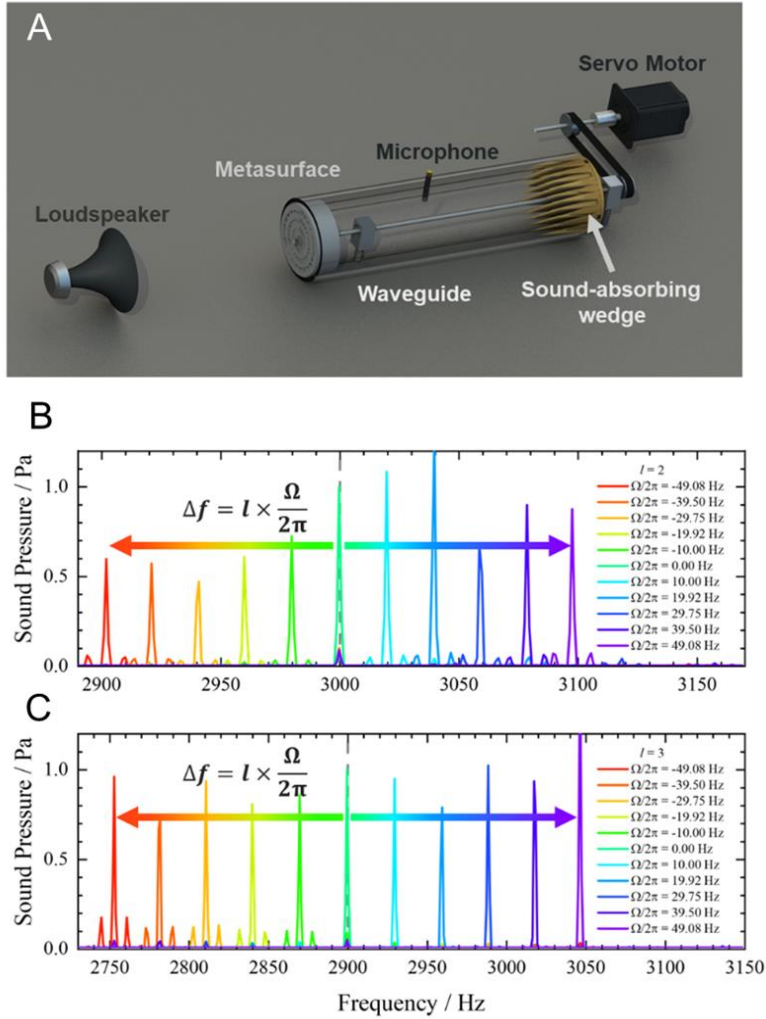

**Figure 3** | (A) Experimental setup for frequency convertor. A metasurface driven by a servo motor is covered by an acoustic waveguide whose end is blocked by a wedge absorber and a microphone is placed 20 cm behind the metasurface. A bearing bracket is used to secure the drive shaft; (B, C) Experimental measurement of the frequency shift under different rotational frequency with topological charge equal to (B) 2 and (C) 3, respectively.

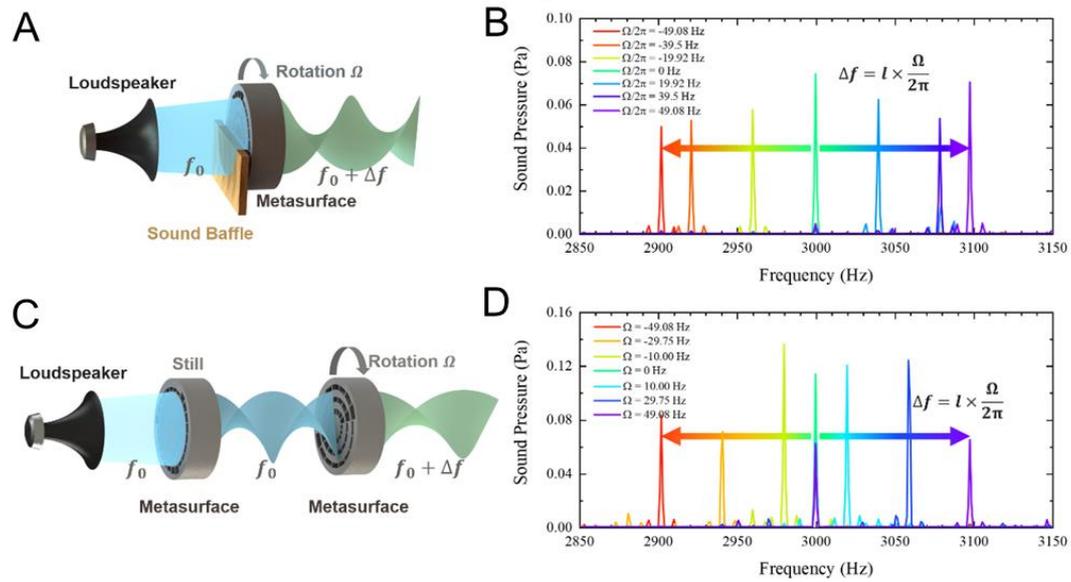

**Figure 4** | (A) Schematic of frequency convertor with the metasurface half covered: the obstacles can be selected arbitrarily; (B) Experimental measurement of the frequency shift under different rotational frequency with topological charge equal to 2 and fundamental frequency equal to 3000 Hz, corresponding to the experimental conditions shown in (A); (C) Schematic of rotational doppler frequency shift with spiral waves incident: a static metasurface is used as a spiral wave emitter; (D) Experimental measurement of the frequency shift under different rotational frequency with topological charge equal to 2 and fundamental frequency equal to 3000 Hz, corresponding to the experimental conditions shown in (C).

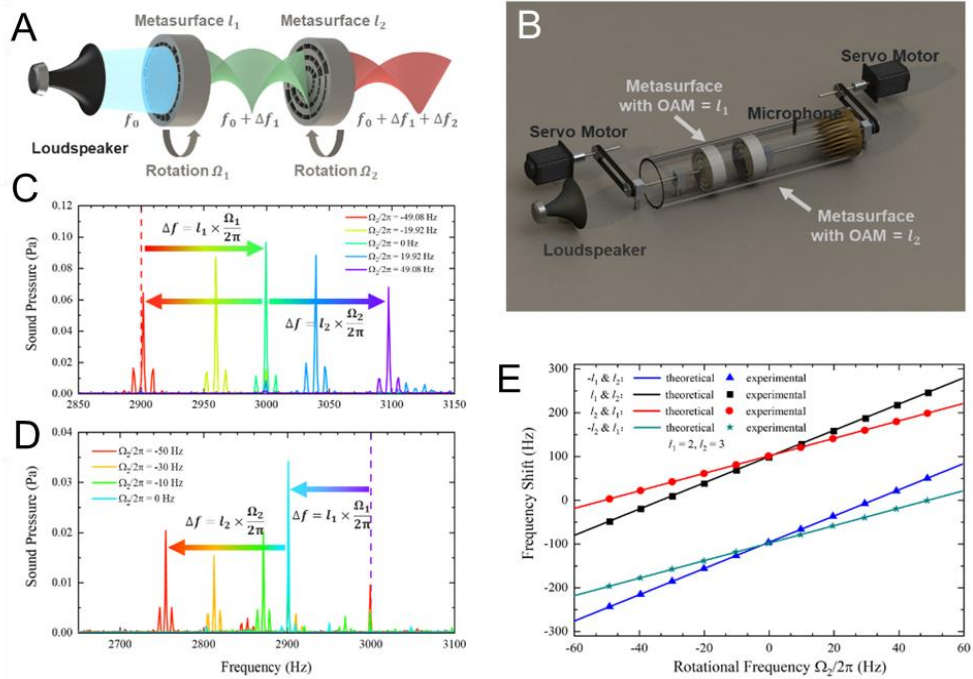

**Figure 5** | (A) Schematic of cascade scheme of the frequency convertor: a metasurface with topological charge equal to $l_1$ rotating at $\Omega_1$ is cascaded with another metasurface topological charge equal to $l_2$ rotating at $\Omega_2$; (B) Schematic of experimental device for the cascade scheme in (A); (C) Experimental measurement of the cascaded frequency conversion under different rotational frequency $\Omega_2$ with $\Omega_1 = 2\pi \times 49.08$ Hz, $l_1 = 2$ and $l_2 = 2$: the fundamental frequency and the working frequency of the metasurface $l_2$ are equal to 2900 Hz and 3000 Hz, respectively; (D) Experimental measurement of the cascaded frequency conversion under different rotational frequency $\Omega_2$ with $\Omega_1 = 2\pi \times 49.08$ Hz, $l_1 = 2$ and $l_2 = 3$: the fundamental frequency and the working frequency of the metasurface $l_2$ are equal to 3000 Hz and 2900 Hz, respectively; (E) Theoretical prediction and experimental measurements of the cascaded frequency shift with different topological charges $l_i$ and rotational frequency $\Omega_i/2\pi$, $i = 1, 2$; $\Omega_2/2\pi = 0$ corresponds to the frequency shift from the metasurface $l_1$ before cascading.